\documentclass[aps,showpacs,amsmath,amssymb,superscriptaddress,twocolumn,prl]{revtex4-1}
\usepackage{graphicx}
\usepackage{CJK}
\usepackage{epstopdf}

\usepackage{amsmath,amssymb}
\usepackage{color}

\newcommand{\wpcm}{W/cm$^2$}

\begin{document}
\begin{CJK*}{UTF8}{gbsn}

\title{Coincidence spectroscopy of high-lying Rydberg states produced in strong laser fields}

\author{Seyedreza Larimian}
\author{Sonia Erattupuzha}
\affiliation{Photonics Institute, Vienna University of Technology, A-1040 Vienna, Austria}
\author{Christoph Lemell}
\author{Shuhei Yoshida}
\author{Stefan Nagele}
\affiliation{Institute for Theoretical Physics, Vienna University of Technology, A-1040 Vienna, Austria}
\author{Raffael Maurer}
\author{Andrius Baltu\v{s}ka}
\affiliation{Photonics Institute, Vienna University of Technology, A-1040 Vienna, Austria}
\author{Joachim Burgd\"orfer}
\affiliation{Institute for Theoretical Physics, Vienna University of Technology, A-1040 Vienna, Austria}
\author{Markus Kitzler}
\author{Xinhua Xie (谢新华)}
\email[Electronic address: ]{xinhua.xie@tuwien.ac.at}
\affiliation{Photonics Institute, Vienna University of Technology, A-1040 Vienna, Austria}

\date{\today}

\begin{abstract}
We report on the measurement of electron emission \textit{after} the interaction of strong laser pulses with atoms and molecules. These electrons originate from high-lying Rydberg states with quantum numbers up to $n \lesssim 120$ formed by frustrated field ionization. Simulations show that both tunneling ionization by a weak dc field and photoionization by the black-body radiation contribute to delayed electron emission on the nano- to microsecond scale. We measured ionization rates from these Rydberg states by coincidence spectroscopy. Further, the dependence of the Rydberg-state production on the ellipticity of the driving laser field proves that such high-lying Rydberg states are populated through electron recapture. The present experiment provides detailed quantitative information on Rydberg production by frustrated field ionization.
\end{abstract}

\maketitle
\end{CJK*}


Ionization of atoms and molecules by strong laser fields is the starting point for a multitude of interesting phenomena, e.g., high harmonic generation or molecular fragmentation \cite{krausz-brabec}. For sufficiently strong laser fields corresponding to intensities of the order of $I\approx 10^{14}$ \wpcm, atoms and molecules are ionized via tunneling ionization, i.e., an electron passes through the potential barrier of the combined Coulomb and laser fields. After tunneling, electrons are steered by the laser field and most of them will eventually escape the Coulomb field of the remaining ion core. However, a fraction of them are recaptured into highly excited states by the ionic Coulomb field. This process frequently referred to frustrated field ionization (FFI) \cite{Nubbemeyer08,Eichmann13} leads to the formation of high-lying Rydberg states with binding energies extending from a fraction of an eV to values of $\mu$eV near threshold.

Very high-lying Rydberg states with principal quantum numbers $n\approx 100$ are quantum objects of macroscopic size allowing for studies of the border between the quantum and the classical worlds \cite{Dunning2009}. Formation and destruction of such mesoscopic objects can be described by semiclassical and classical methods \cite{gallagher2005rydberg}. Recent experiments on high harmonic generation and electron wave packet interferometry indicate the important contribution of such excited states to different processes \cite{chini2014coherent,xie12,Deng2015,Xie2015,arbo14} including ionization and molecular dissociation processes \cite{Wolter14,li14,liu12,minns14}. However, detailed and quantitative information on the FFI following the interaction of femtosecond laser pulses with atoms and molecules appears to be scarce.

To explore the production process and the properties of high-lying Rydberg states formed in the strong field interaction with atoms and molecules, direct observation of such states is required. Traditionally, zero kinetic energy photoelectron spectroscopy is applied to study weakly bound states in atoms and molecules \cite{zeke}. In case of strong field interaction, however, the ionization signal from Rydberg states is completely overshadowed by the dominant laser field-induced ionization signal from the target and the residual gas in the interaction chamber. Therefore, the signal from Rydberg states cannot be extracted easily.

Signatures of post-pulse ionization of high-lying Rydberg states have been found recently in momentum spectra of photoelectrons (``zero-energy structures'', \cite{Wolter14,liu12}). In this Letter, we report on the experimental observation of high-lying Rydberg states with principal quantum number up to $n\lesssim 120$ populated during strong field interaction of atoms and molecules. Using a reaction microscope \cite{ullrich03,doerner00} we perform coincidence measurements of electrons and ions separated from each other by the ionization process. The delayed signal of weak dc field induced ionization of high-lying Rydberg states can be well distinguished from the prompt strong field ionization signal and retains a very high signal-to-noise ratio. We identify two characteristic time constants governing the delayed ionization of Rydberg states: (i) a strongly dc field dependent and (ii) a nearly field independent contribution attributed to emission from long-lived Stark resonances and photoionization by black-body radiation (BBR) at room temperature \cite{gallagher1979interactions}. We also determine the dependence of the strong-field induced Rydberg population on the ellipticity of the exciting laser pulse.


Measurements were carried out with laser pulses from a home-built Ti:Saphire laser amplifier with a center wavelength of 795 nm, a repetition rate of 5 kHz and a pulse duration (full width at half maximum of the intensity) of 25 fs. The peak laser intensity is on the order of $I \sim 10^{14}$ \wpcm\ with a peak electric field strength of $F_0\sim 2.7\times 10^{8}$ V/cm. A weak homogeneous dc field from $F_\mathrm{dc}=0.1$ V/cm up to $F_\mathrm{dc}=30$ V/cm is applied along $z$-direction in the time-of-flight (TOF) spectrometer to accelerate charged particles to the detectors but also to induce static field ionization of high-lying Rydberg states populated during the strong field interaction of atoms and molecules. Additionally, a homogeneous magnetic field of 12.3 gauss ensures 4$\pi$ detection of electrons. More details on the experimental setup can be found in, e.g., \cite{xie12,xie12_2}. We performed the measurements with three atomic gases (argon, helium, and neon) and six molecular targets (hydrogen, methane, ethylene, acetylene, 1,3-butadiene, and hexane). We observe field induced ionization of high-lying Rydberg states for all three atomic gases and for hydrogen, methane, ethylene, and acetylene molecular targets but not for 1,3-butadiene and hexane. This may be related to the instability of high-lying Rydberg states of such large molecules leading to molecular dissociation on time scales shorter than the delayed detection of Rydberg states starting at a few hundred nanoseconds~\cite{Zhou2012,Bogomolov2014}. In the following we quantitatively analyze coincidence spectra of post-pulse ionized Rydberg states of argon. We find for these Rydberg states delayed emission times extending to microseconds. Similar results were found for all other target species for which this process was observed.


\begin{figure}[htbp]
\centering
\includegraphics[width=0.45\textwidth]{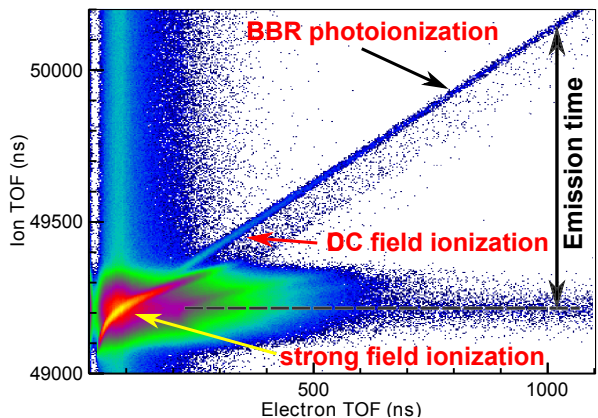}
\caption{(Color online) PEPICO spectrum for argon atoms interacting with 25 fs laser pulses with peak electric field strength of $F_0=3.4\times10^8$ V/cm (intensity $I=1.5\times10^{14}$ \wpcm) and a dc field strength of $F_\mathrm{dc}=3$ V/cm. Signals of correlated electrons and ions are found along the diagonal. Electron emission during and after the laser pulse can be distinguished in this spectrum.} \label{fig:pepico}
\end{figure}
A typical photoelectron-photoion-coincidence (PEPICO) spectrum for argon atoms interacting with a 25 fs laser pulse ($I=1.5\times10^{14}$ \wpcm, $F_\mathrm{dc}=3$ V/cm) features a main peak at electron TOF of 85 ns and ion TOF of 49205 ns (Fig.~\ref{fig:pepico}) representing photoemission in the strong field of the laser pulse and setting the reference for the delayed emission. Post-pulse ionization events, i.e., an electron and its parent ion being separated after the conclusion of the exciting laser pulse with initial momentum $p_{e,i}\approx 0$ a.u.\ are registered along the diagonal with unit slope from which the ionization times of the excited argon atoms can be read out directly. Clear signals for post-pulse ionization were found for emission times extending up to 80 microseconds after conclusion of the laser pulse. Our interpretation of the PEPICO spectrum is confirmed by trajectory simulations to determine the times of flight of atoms photoionized by strong field interaction and from post-pulse ionization by the dc field reproducing the parabolic main feature and the diagonal line, respectively.

\begin{figure}[htbp]
\centering
\includegraphics[width=0.45\textwidth]{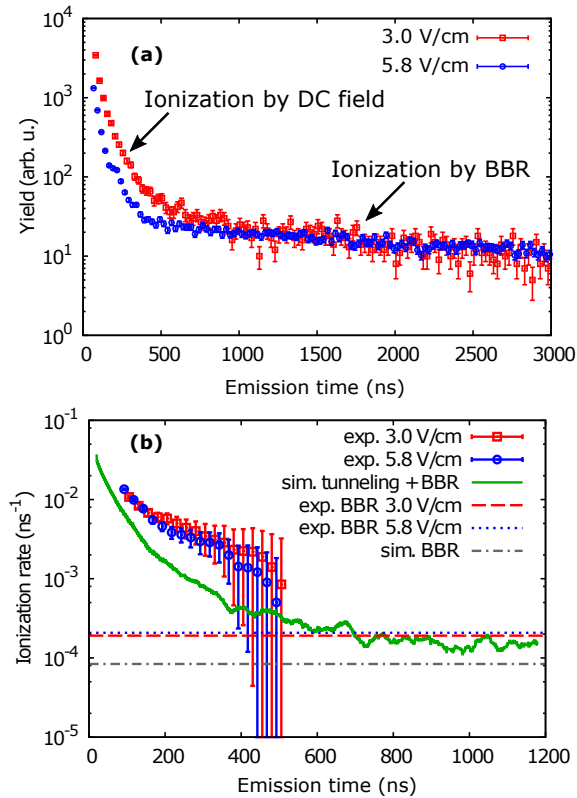}
\caption{(Color online) (a) Intensity $I(\tau)$ of the signal of post-pulse ionized Rydberg states of argon (diagonal in Fig.~\ref{fig:pepico}) as a function of the emission time $\tau$ for two field strengths $F_\mathrm{dc}$ with the same laser interaction condition (peak intensity of $1.5\times 10^{14}$\wpcm). (b) Ionization rate of Rydberg states derived from the dada in (a) in comparison with simulated tunneling ionization rates for $F_\mathrm{dc}=3$ V/cm and black-body radiation (BBR) rates.}
\label{fig:lifetime}
\end{figure}

When ions recapture electrons released during the laser pulse, high-$n$ states of the atom with $n\gtrsim n_\alpha=\sqrt{\alpha}$ are populated \cite{Wolter14,Arbo15} with $\alpha=F_0/\omega_\mathrm{ir}^2$ the quiver amplitude (atomic units are used throughout unless otherwise noted). For the laser parameters of Fig.\ \ref{fig:pepico} $n_\alpha\approx 12$ and the dominant angular momenta of these states have been found to be close to $L\approx \sqrt{2 F_0}/\omega=6$ \cite{Arbo15}. In the presence of the weak dc field $F_\mathrm{dc}$, Rydberg states very close to the continuum threshold and well above the potential barrier can be ionized. Due to ``fast'' laser excitation compared to ``slow'' dynamics of Rydberg electrons the lowest $n_F$ Rydberg state being ionized can be estimated using the diabatic field ionization threshold $F_\mathrm{dc} = 1/(9 n_F^4)$. For $F_\mathrm{dc}=3$ V/cm Rydberg Stark states with $n > n_F \simeq 120$ are accessible to post-pulse over-the-barrier ionization. Such states have a diameter of more than 0.5 $\mu$m ($\langle r\rangle_n\propto n^2$) and typical orbital periods $\tau_\mathrm{orb}= 2\pi n^3$ of a hundred picoseconds. The emission time from strong-field induced Rydberg population can be derived from the intensity distribution along the diagonal in Fig.\ \ref{fig:pepico} [plotted in Fig.\ \ref{fig:lifetime}(a)]. For emission times smaller than $\tau< 100$ ns a clear distinction between strong-field and post-pulse ionization was not possible within the resolution of our experiment.

The post-pulse ionized yield $I(\tau)$ strongly depends on the field strength $F_\mathrm{dc}$ of the external dc field [Fig.\ \ref{fig:lifetime}(a)]. As expected, stronger dc fields initially lead to faster field ionization of the excited Ar$^\ast$ atoms. By contrast, the corresponding ionization rates $\Gamma=-d\ln[I(\tau)]/d\tau$ [Fig.\ \ref{fig:lifetime}(b)] derived from the data in Fig.\ \ref{fig:lifetime}(a) appear to be rather insensitive to the value of $F_\mathrm{dc}$. The rate decreases from about $\Gamma\approx 0.01$ ns$^{-1}$ at $\tau=100$ ns to about $\Gamma\approx 0.001$ ns$^{-1}$ at $\tau=500$ ns. For emission times longer than 650 ns the ionization rate becomes nearly constant at $\Gamma\approx 2 \times 10^{-4}$ ns$^{-1}$.
The constant ionization rates for the two field strengths, indicated by blue dotted (at $2.08\times 10^{-4}$ ns$^{-1}$) and red dashed (at $1.91\times 10^{-4}$ ns$^{-1}$) horizontal lines in Fig.~\ref{fig:lifetime}(b), are derived from the measured signals in Fig.~\ref{fig:lifetime}(a) for emission times larger than 650 ns by fitting with exponential functions.
For comparison, the strong field tunneling ionization rate for the argon ground state ($E_\mathrm{bind}=-15.76$ eV) for a field strength of $F_0=3.4\times 10^8$ V/cm (intensity $I=1.5\times10^{14}$ \wpcm) is, according to the ADK formula \cite{ADK}, about $ 2 \times 10^5$ ns$^{-1}$, i.e., many orders of magnitude larger.

\begin{figure}[htbp]
\centering
\includegraphics[width=0.45\textwidth]{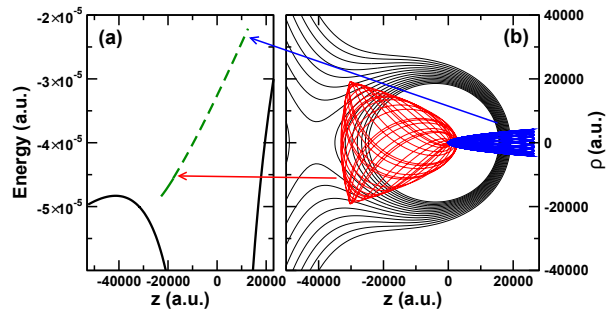}
\caption{(Color online) (a) Potential energy along $z$ axis and eigenenergies of hydrogenic Stark states with $n=121, m=0$ in a weak dc field of $F_\mathrm{dc}=3$ V/cm. The energy is plotted as a function of the expectation value $\langle z\rangle$ of the associated Stark state. The solid part of the line indicates states with tunneling rate $> 10^{-7}$~ns$^{-1}$, the dashed part states with smaller rate. (b) Typical classical trajectories of red- and blue-shifted Stark states with corresponding energies indicated by arrows. Contour lines of the potential landscape are shown as thin black lines.} \label{fig:peda}
\end{figure}

In order to identify processes underlying the delayed emission on the nano- to microsecond scale, different ionization channels must be considered: over-the-barrier and tunneling ionization, and black-body radiation induced photoionization and photoexcitation to above-barrier states. For a field strength $F_\mathrm{dc}=3$ V/cm which we focus on in the following the saddle of the potential landscape lies energetically at $n=102$ of the unperturbed hydrogenic spectrum. In the dc field the critical primary quantum number is shifted to $n_F= 121$ (Fig.\ \ref{fig:peda}). For states lying well below the Stark barrier, tunneling rates are negligibly small on the nano- to microsecond timescale the experiment is sensitive to. Stark states above the saddle form long-lived shape resonances. The corresponding classical trajectories are bound [Fig.\ \ref{fig:peda}(b)]. The blue-shifted state of the $n=121$ manifold points in the direction opposite to the saddle and features lifetimes long compared to the ionization accessible in the experiment. Even most of the red-shifted states localized on the downhill side miss the saddle because of the relatively large transverse energy $p_\perp^2/2$. Quantum mechanically, the latter have tunneling rates $> 10^{-7}$ ns$^{-1}$ (corresponding to quantum numbers $n=121,\, n_1<10$, and $m=0$) indicated with a solid line in the energy diagram in Fig.\ \ref{fig:peda}(a) while states with larger $n_1$ have ionization rates $< 10^{-7}$ ns$^{-1}$ [dashed part of the line in Fig.\ \ref{fig:peda}(a)]. The applicability of hydrogenic Stark ionization rates to the present data for strong-field ionization of argon follows from the fact that frustrated field ionization forms predominantly non-core penetrating high-$\ell$ states which are well approximated by hydrogenic states. Scattering at the non-hydrogenic core of argon could transfer initially blue-shifted Stark states to states oriented towards the saddle leading to non-negligible contributions of tunneling ionization \cite{Spencer1982,Beterov2007}. However, in the presence of the dc field the core scattering
rate is strongly suppressed and too small to efficiently depopulate the long-lived Stark resonances during the accessible time interval.
On the (sub-)microsecond time scale ionization by BBR present in the experiment performed at room temperature,
$\langle \omega_\mathrm{BBR}\rangle\approx 0.025$~eV, become non-negligible. BBR may either directly ionize the atom or excite it to even higher Rydberg states eventually leading to over-the-barrier or tunneling ionization~\cite{Spencer1982,Beterov2007}.

The time-dependent Rydberg population of $m=0$ states with parabolic quantum number $n_1$, $\rho(n,n_1,t)$, is governed by the rate equation
\begin{equation}
\dot\rho(n,n_1,t) = -\left[T(n,n_1)+R_\mathrm{BBR}(n)\right]\rho(n,n_1,t)\, ,\label{re1}
\end{equation}
with $T(n,n_1)$ the tunneling ionization rate and $R_\mathrm{BBR}(n)$ the black-body radiation photoionization rate. We determine the effective time-dependent ionization rate
\begin{equation}
\Gamma_\mathrm{eff}(t)=-\frac{d\ln\left(\sum_{n,n_1}\rho(n,n_1,t)\right)}{dt}
\end{equation}
by Monte-carlo sampling of decay channels starting with an energy-dependent spectral excitation density $\rho(n,0) dn/dE\approx c_0$ formed by frustrated field ionization \citep{Wolter14}. The such obtained ionization rate for Rydberg states with $n=121$ to $n=136$ are independent of $n$ (within statistical error). Therefore, the calculated rates are rather insensitive to the $n$-distribution of Rydberg states after laser excitation.
Results for $n=121$ are shown in Fig.\ \ref{fig:lifetime}(c).
The initial rapid decay of the ionization rate within 500 ns can be attributed to tunneling of red-shifted Stark states.

The black-body radiative ionization rate $R_\mathrm{BBR}(n)$, while overall very small, increases with decreasing $n$ down to quantum numbers $n_\mathrm{BBR}$ where the binding energy matches $\langle\omega_\mathrm{BBR}\rangle$ \cite{Spencer1982,Beterov2007}. In the present case $n_\mathrm{BBR}\approx 20$. Summing over the range $n < n_F$ the BBR ionization rate is estimated as $R_\mathrm{BBR}(n)=8.4 \times 10^{-5}$~ns$^{-1}$. Additionally the BBR-induced excitation rate is about $1\times 10^{-5}$~ns$^{-1}$. These contributions are comparable to the ionization rate from tunneling. Total ionization rates $\sim 2 \times 10^{-4}$~ns$^{-1}$ including tunneling and BBR-induced
ionization yield good agreement with the observed time dependence of the delayed emission for times approaching one microsecond. Small differences can be attributed to the effect of stray fields and inhomogeneities of the field.

The simulation suggests that the presented time-delayed PEPICO spectra provide for the first time direct access to long-lived Stark resonances in very high Rydberg states and to the population of the Rydberg manifolds extending from $n_\mathrm{BBR}\approx 20$ to $n_F\approx 120$ formed by frustrated field ionization.

\begin{figure}[htbp]
\centering
\includegraphics[width=0.45\textwidth]{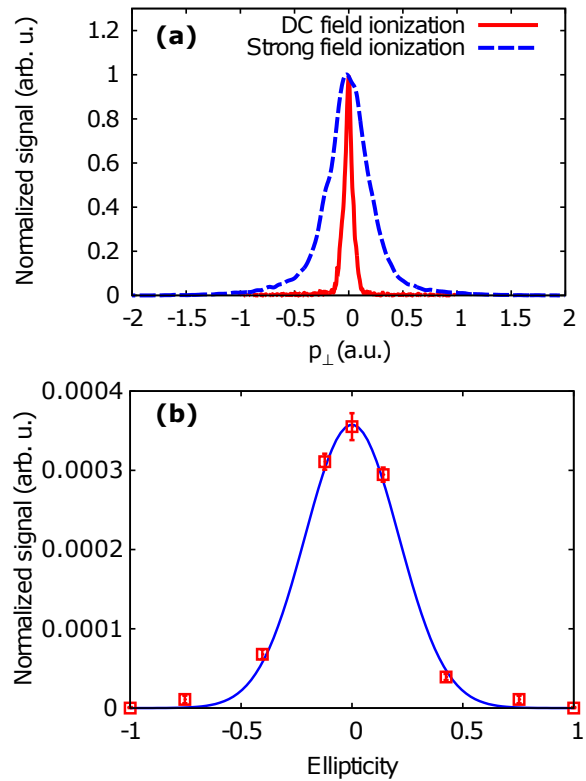}
\caption{(Color online) (a) Projected momentum distribution of delayed electron emission perpendicular to the direction of $F_\mathrm{dc}$ (the red solid line) and of prompt electrons perpendicular to the laser polarization direction (the blue dashed lines).
(b) Measured time-integrated signal of delayed ionization of high-lying Rydberg states of Ar normalized to the strong laser field ionized Ar as a function of laser ellipticity (red points). The blue solid line is to guide the eye.} \label{fig:dependence}
\end{figure}

With our reaction microscope we also measure the initial momenta of the released Rydberg electrons. As expected for a ``zero energy structure'', the average momentum of the post-pulse ionized electrons is very small. Moreover, we find the momentum distribution perpendicular to the direction of the dc field ($\Delta p_\perp^\mathrm{dc}=0.04$ a.u.) to be much narrower than the transverse distribution of the prompt electron emission by strong-field ionization $\Delta p_\perp=0.4$ a.u.\ (Fig.\ \ref{fig:dependence}).
For strong field emission the width of the momentum distribution is given by
\begin{equation}
T(n,p_\perp) = C\exp\left(-\frac{2}{3n^3F}\right)\exp\left(-\frac{p_\perp^2}{nF}\right)\label{eq2}
\end{equation}
and has been previously derived \cite{Arissian2010} to give $\Delta p_\perp=\sqrt{2\log 2\cdot F/|E_\mathrm{bind}|}=0.41$ a.u.\ in close agreement with our data (Fig.\ \ref{fig:dependence}). Applying the same expression to high Rydberg states $n\approx 100$ yields $\Delta p_\perp^\mathrm{dc}\approx 0.004$ a.u., much narrower than the experimental momentum resolution which is about 0.04 a.u. Likewise, BBR ionized electrons are expected to have a momentum distribution peaked near zero momentum, $|p|\ll \sqrt{2\langle\omega_\mathrm{BBR}\rangle}\approx 0.04$. Therefore, the width of the delayed peak in Fig.~\ref{fig:dependence} is primarily given by our experimental resolution.

The ellipticity of the laser pulse is expected to have a strong influence on the probability for recapturing electrons after tunneling and, hence, on $\rho (n,0)$. In our experiment, we have tested the dependence of the Rydberg signal on the laser ellipticity with a laser intensity of $I=2.5\times 10^{14}$ \wpcm\ and a dc field strength of $F_\mathrm{dc}=3$ V/cm [Fig.\ \ref{fig:dependence}(b)].
The Rydberg signal has a maximum for linearly polarized laser fields and drops to 0 when increasing the ellipticity up to 1 (circular polarization). This ellipticity dependence is similar to that of high harmonic generation from strong laser interaction \cite{Liang1995,Flettner2002,Moller2012} and to the ellipticity dependence of the occupation of Rydberg states with $n<30$ in helium \cite{Nubbemeyer08,Landsman2013}. The observed ellipticity dependence confirms that the formation of the high-lying Rydberg states is a rescattering process with the released electron ending up close to the ionic core with small momentum at the conclusion of the pulse. In laser pulses with ellipticity $\varepsilon\neq 0$ the electron is driven away from the core effectively suppressing the recapture process.

In conclusion, we have measured the post-pulse delayed ionization of high-lying Rydberg states populated during the interaction of intense laser pulses with atoms and molecules in the presence of a weak electric dc field using coincidence spectroscopy. Recapture of tunnel ionized electrons, frequently referred to as frustrated field ionization, appears to be a very general process as we have observed its signatures for various atomic and molecular target species. The time-delayed coincidence spectroscopy allows to identify two major contributions: tunneling ionization of Rydberg Stark resonances lying just above the saddle point of the potential landscape in the presence of a weak dc field, and photoionization by black-body radiation of lower lying Rydberg states. A gradual transition from the tunneling dominated regime was observed for the field strength $F_\mathrm{dc}=3$ V/cm near $\tau\approx 500$ ns. Tuning the strength of the external dc field strength $F_\mathrm{dc}$ may open the pathway to selectively probe Stark resonances close to the principal quantum number $n_F\lesssim 3^{-1/2} F_\mathrm{dc}^{-1/4}$ near the saddle threshold for over-the-barrier ionization. The experimental data are consistent with a nearly constant spectral excitation density of the Rydberg manifold above $n\gtrsim 20$. We also find that frustrated field ionization is strongly dependent on the ellipticity of the driving laser pulse.

We would like to thank Armin Scrinzi and Chii-Dong Lin for fruitful discussions.
This work was supported by the Austrian Science Fund (FWF): P25615-N27, P21141-N16, P28475-N27, P21463-N22, P27491-N27, special research programs SFB-041 ViCoM, SFB-049 Next Lite and doctoral college W1243. Calculations were performed using the Vienna Scientific Cluster (VSC).

\end{document}